\documentclass{appolb}
\usepackage{epsfig}
\usepackage{cite}

\begin{document}
\pagestyle{plain}
\eqsec
\newcount\eLiNe\eLiNe=\inputlineno\advance\eLiNe by -1
\title{\bf {Next-to-leading order QCD corrections to $t\bar{t}b\bar{b}$ \\
production at the LHC} 
\thanks{Presented at the XXXIII International Conference of Theoretical 
Physics, Matter To The Deepest: Recent Developments In Physics of Fundamental 
Interactions, Ustro\'n, 11-16 September 2009, Poland.}
\thanks{Preprint number: WUB/09-12}
}
\author{Ma\l gorzata Worek
\address{Fachbereich C Physik, Bergische Universit\"at Wuppertal\\
D-42097 Wuppertal, Germany \\
\vspace{0.2cm}
e-mail: \texttt{worek@physik.uni-wuppertal.de}
}}
\maketitle

\begin{abstract}
In this contribution, we summarize the results from an NLO computation of
the production of $t\bar{t}b\bar{b}$ in hadronic collisions. The results are
obtained by combining two programs: \textsc{Helac-1Loop}, based on the OPP
method and the reduction code \textsc{CutTools}, for the virtual  one-loop
matrix elements and \textsc{Helac-Dipoles} for the real-emission
contributions. Selected numerical results are presented for the LHC. 
\end{abstract}

\section{Introduction}

The process $pp \rightarrow t\bar{t}b\bar{b}$ represents a very important
background reaction to searches at the LHC, in particular to $t\bar{t}H$
production, where the Higgs boson decays into a $b\bar{b}$ pair. A successful
analysis of this  production channel at the LHC requires the  knowledge of
direct  $t\bar{t}b\bar{b}$ production at NLO in  QCD
\cite{Aad:2009wy}. Moreover, the calculation of NLO QCD  corrections to $2
\rightarrow 4$ processes at the LHC represents the current  technical
frontier. The complexity of such calculations triggered the  creation of
prioritized experimenters wishlists where the  $t\bar{t}b\bar{b}$ production
ranges among the most  wanted candidates \cite{Bern:2008ef}. The NLO
computation for this process  has been completed only very recently in
\cite{Bredenstein:2009aj} and  then confirmed in \cite{Bevilacqua:2009zn}
where we demonstrated the power of our system based on 
\textsc{Helac-Phegas}\footnote{\tt 
http://helac-phegas.web.cern.ch/helac-phegas/}
\cite{Kanaki:2000ey,Papadopoulos:2000tt,Cafarella:2007pc},  which has, on its
own, already been extensively used and tested in phenomenological studies see
\eg
\cite{Gleisberg:2003bi,Papadopoulos:2005ky,Alwall:2007fs,Englert:2008tn},
\textsc{Helac-1Loop} \cite{vanHameren:2009dr}, which is not yet publicly
available,   
\textsc{CutTools}\footnote{\tt http://www.ugr.es/$^{\sim}$pittau/CutTools/}
\cite{Ossola:2007ax}  and \textsc{Helac-Dipoles}\footnote{\tt 
http://helac-phegas.web.cern.ch/helac-phegas/}
\cite{Czakon:2009ss} in a realistic computation with six external legs and
massive partons.  In the following we briefly summarize the calculation of
the virtual and real corrections and present numerical results for the LHC.

\section{Virtual corrections}

A one-loop $n-$particle amplitude can be expressed in terms of a basis of
known $4-$, $3-$, $2-$ and  $1-$point scalar integrals: boxes, triangles,
bubbles and tadpoles. The coefficients depend in general on the dimension of
space-time, $d$, which, upon expansion around $d=4$, generates a rational
function in the invariants (rational term). The concept behind modern methods
of evaluation, is the direct determination of the coefficients, without
recurring to a Passarino-Veltman reduction.

In our case, the coefficients and one part of the rational term (see
\cite{Ossola:2008xq}) are extracted via the OPP reduction method at the
integrand level \cite{Ossola:2006us}, which is implemented  in
\textsc{CutTools}. The second part of the rational term coming from 
the epsilon-dimensional contributions in the numerator is computed with the
help of dedicated Feynman rules \cite{Draggiotis:2009yb,Garzelli:2009is}. 

The OPP reduction is based on a representation of the numerator of amplitudes, 
a polynomial in the integration momentum, in a basis of polynomials given by
products of the functions in the denominators. Clearly, the cancellation of
such terms with the actual denominators will lead to scalar functions with a
lower number of denominators. By virtue of the proof provided by the
Passarino-Veltman reduction, we will end up with a tower of four-point and
lower functions, as mentioned before. The determination of the decomposition
in the new basis proceeds recursively, by setting chosen denominators
on-shell. This is where the OPP method resembles generalized 
unitarity \cite{Bern:1994zx,Bern:1994cg,Witten:2003nn,Britto:2004nc,
Bern:2007dw,Ellis:2007br,Giele:2008ve}. 
For most recent applications see \eg \cite{Ellis:2009zw,Berger:2009zg,
KeithEllis:2009bu,Berger:2009ep,Melnikov:2009dn,Melnikov:2009wh}

It is important to stress, that working around four dimensions, allows to
compute the numerator function in four dimensions. The difference to the
complete result is of the order of $\epsilon$, and can therefore be 
determined a posteriori in a simplified manner. Since the calculation of the
coefficients of the reduction requires the evaluation of the numerator
function for a given value of the loop momentum, the corresponding diagrams
can be thought as tree level (all momenta are fixed) graphs. To complete the
analogy, one needs to chose a propagator and consider it as cut. At this point
the original amplitude for an $n$ particle process becomes a tree level
amplitude for an $n+2$ particle process. The advantage is that its value can be
obtained by a tree level automate such as \textsc{Helac}. The bookkeeping 
necessary for a practical implementation is managed by a new software, 
\textsc{Helac-1Loop}.

To recapitulate, a complete calculation involves three building components:
1) \textsc{CutTools}, for the reduction of integrals with a given numerator 
to a basis of scalar functions; 2) \textsc{Helac-1Loop} for the evaluation of 
the numerator functions for given loop momentum (fixed by \textsc{CutTools}); 
3) \textsc{OneLOop}\footnote{The package includes  all divergent and finite 
scalar integrals, with massless and massive propagators, and with the UV as 
well as the IR divergencies dealt with within dimensional regularization. It 
can be obtained from {\tt http://annapurna.ifj.edu.pl/$^{\sim}$hameren/}} 
\cite{vanHameren:2009dr}, a library of scalar functions, which provides the 
actual numerical values of the integrals. At the end, the 1-loop result is
given in the form of real and imaginary parts of the finite term and of the
coefficients of the $1/\epsilon$ and $1/\epsilon^2$ poles. 

The above procedure provides the bare 1-loop amplitude. Renormalization is
performed as usual, by evaluating tree level diagrams with
counterterms. For our process, we chose to renormalize the coupling in the
$\overline{\rm {MS}}$ scheme, but the mass in the on-shell scheme (wave function
renormalization is done in the on-shell scheme as it must be). Notice, that
also this part is performed in four dimensions. This means, that the whole
procedure is equivalent to the 't Hooft-Veltman version of dimensional
regularization \cite{'tHooft:1972fi}.

Let us emphasise that all parts are calculated fully numerically  
in a completely automatic manner. 

\section{Real corrections}

The singularities from soft or collinear gluon emission are isolated via
dipole subtraction for NLO QCD calculations \cite{Catani:1996vz} using the
formulation for massive quarks \cite{Catani:2002hc} and for arbitrary
polarizations \cite{Czakon:2009ss}. After combining virtual and real
corrections, singularities connected to collinear configurations in the final
state as well as soft divergencies in the initial and final states cancel for
collinear-safe observables automatically after applying a jet
algorithm. Singularities connected to collinear initial-state splittings are
removed via factorization by PDF redefinitions.   We do not use finite dipoles
regularizing the quasi-collinear divergence induced by both top quarks moving
in the same direction. Due to the large top quark mass, they are not needed to
improve the numerical convergence. 

Calculations are performed with the help of the \textsc{Helac-Dipoles} software,
which is a complete and publicly available automatic implementation  of
Catani-Seymour dipole subtraction and consists of  phase space integration  of
subtracted real radiation and integrated dipoles in both massless and massive
cases.  Let us stress at this point, that a phase space restriction on the
contribution of the dipoles as originally proposed in
\cite{Nagy:1998bb,Nagy:2003tz} is also implemented. All results presented in
the next section have been obtained with $\alpha_{max}=0.01$ parameter.   The
phase-space integration is performed with the multichannel Monte Carlo  generator \textsc{Phegas} \cite{Papadopoulos:2000tt} and optimized   via
\textsc{Parni}\footnote{\tt http://annapurna.ifj.edu.pl/$^{\sim}$hameren/}
\cite{vanHameren:2007pt}.

\section{Numerical results}

We present predictions for $pp \rightarrow  t\bar{t}b\bar{b} + X$ at $\sqrt{s}
= 14$ TeV. For the top-quark mass we take $m_t = 172.6$ GeV,  while all other
QCD partons, including b quarks, are treated as massless.  Final states
involving collinear gluons and b-quarks are recombined into  collinear-safe
jets by means of the $k_T$-algorithm
\cite{Catani:1992zp,Catani:1993hr,Ellis:1993tq}. Specifically we require two
b-quark jets with separation  $\sqrt{\Delta\phi^2 +\Delta y^2} > D = 0.8$ in
the rapidity– azimuthal-angle plane. Motivated by the search for a
$t\bar{t}H(H \rightarrow b\bar{b})$ signal at the LHC \cite{Aad:2009wy},  we
impose the following additional cuts on the transverse  momenta and rapidity
of the b-quark jets: $p_{T} > 20$ GeV, $|y| < 2.5$.  The outgoing (anti)top
quarks are neither affected by the jet algorithm nor by phase-space cuts.  We
use CTEQ6 PDFs \cite{Pumplin:2002vw,Stump:2003yu}. Most precisely 
we use the CTEQ6M parton distributions at NLO, and the CTEQ6L1 set at LO.
The suppressed contribution from b quarks in the initial state has been 
neglected.  

\begin{figure}[th]
\begin{center}
\includegraphics[width=0.55\textwidth]{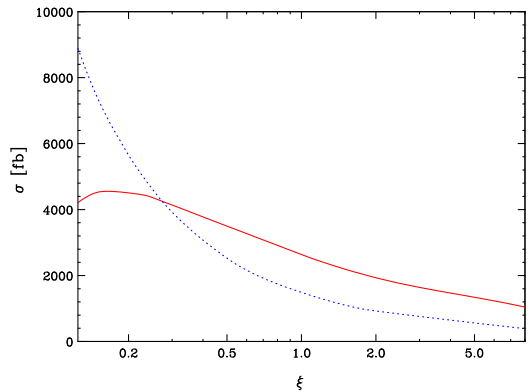}
\end{center}
\vspace{-0.2cm}
\caption{\it \label{fig:scale} Scale dependence of the total cross
section for $pp\rightarrow  t\bar{t}b\bar{b} + X$ at the LHC  
with $\mu_R=\mu_F=\xi m_t$. The blue dashed curve corresponds to 
the leading order, whereas the red solid one to the next-to-leading 
order result. }
\end{figure}
\begin{figure}[th]
\begin{center}
\includegraphics[width=0.45\textwidth]{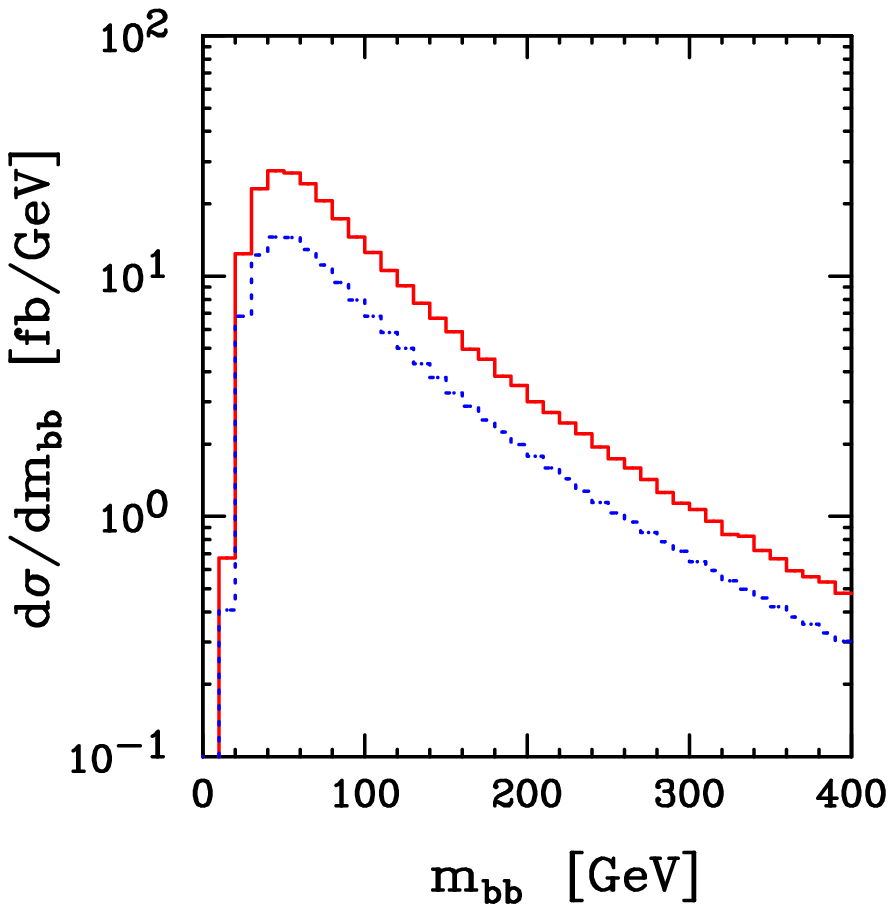}
\includegraphics[width=0.425\textwidth]{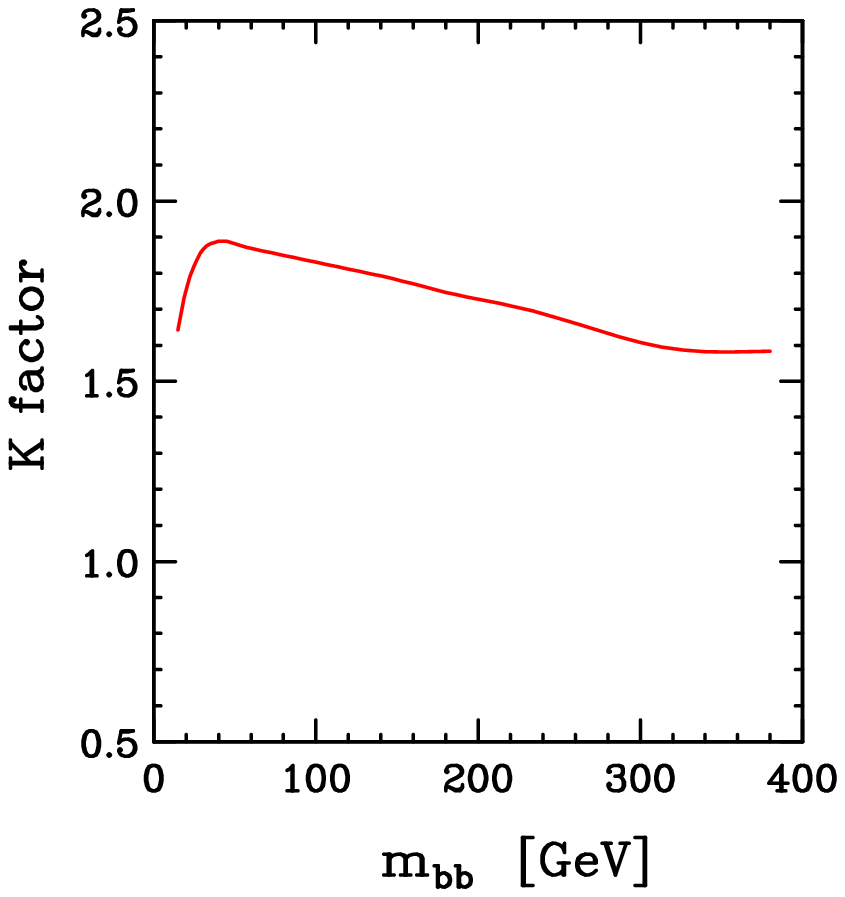}
\end{center}
\vspace{-0.2cm}
\caption{\it \label{fig:mbb} Right panel: distribution of 
the invariant mass $m_{b\bar{b}}$ of the $b\bar{b}$ pair for 
$pp\rightarrow t\bar{t}b\bar{b} + X$ at the LHC at LO 
(blue dashed line) and NLO (red solid line). 
Left panel: ratio of the NLO and LO distributions.}
\end{figure}
\begin{figure}[t]
\begin{center}
\includegraphics[width=0.45\textwidth]{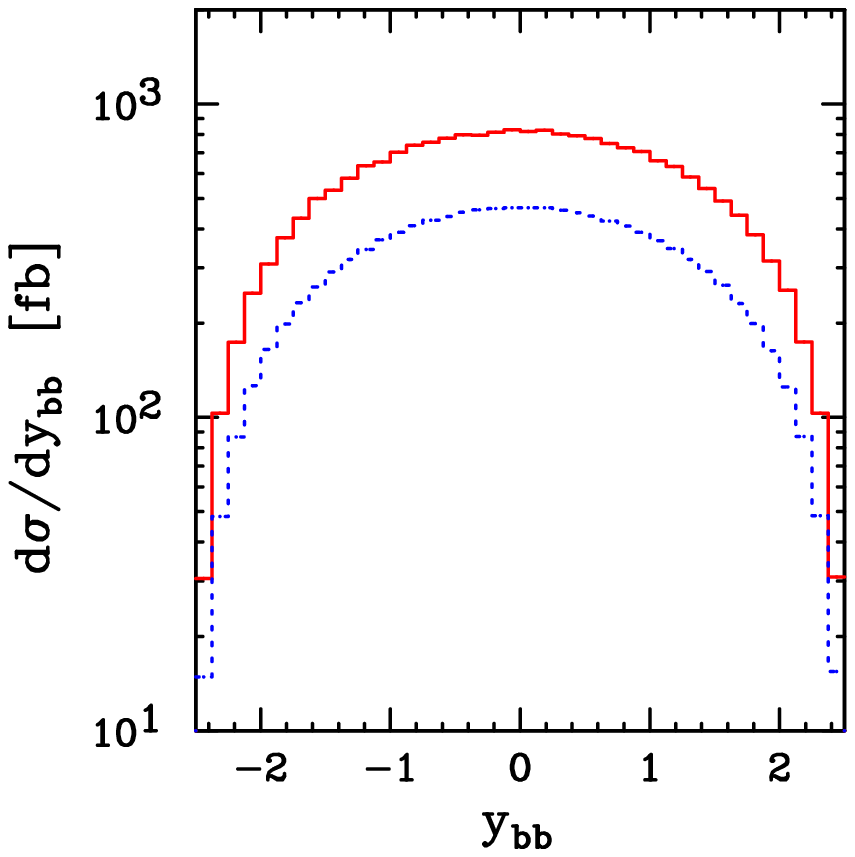}
\includegraphics[width=0.425\textwidth]{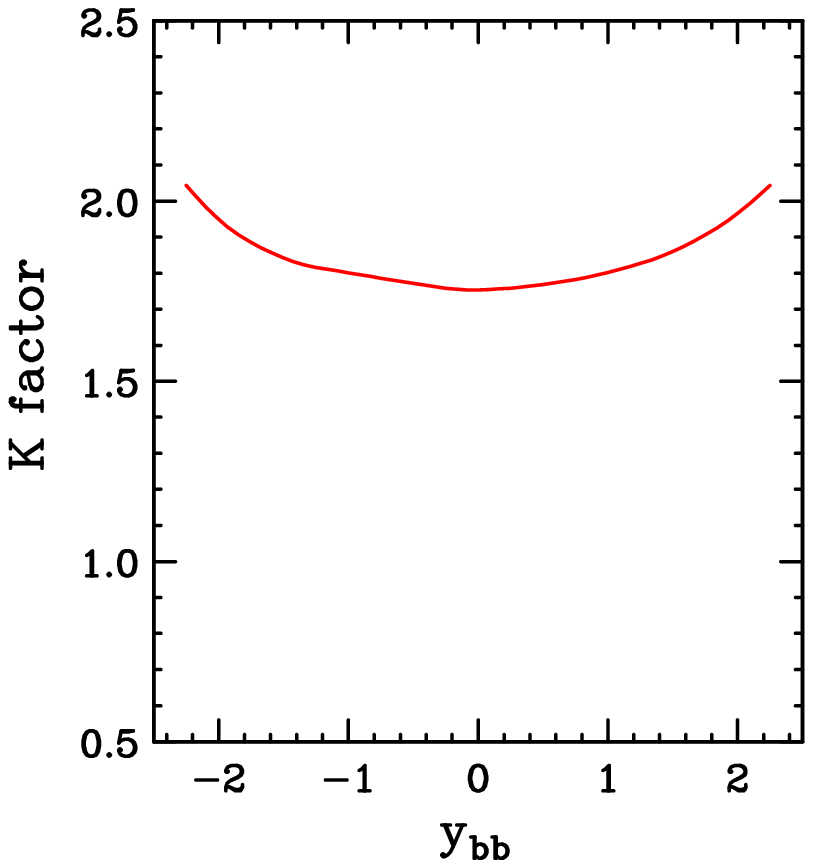}
\end{center}
\vspace{-0.2cm}
\caption{\it \label{fig:ybb} Right panel: distribution in the rapidity 
$y_{b\bar{b}}$ of the $b\bar{b}$ pair for $pp\rightarrow t\bar{t}b\bar{b}+ X$ 
at the LHC at LO (blue dashed line) and NLO (red solid line). 
Left panel: ratio of the NLO and LO distributions.}
\end{figure}

We start with a presentation of the total cross section  at the central value
of the scale, $\mu_R=\mu_F=m_t$ at LO and NLO.  At the central scale, the full
pp cross section receives a very large NLO  correction of the order of 
77\%, which is
mainly due to the gluonic initial states. The full LO and NLO cross sections
are given by $\sigma_{LO}=1489.2$ fb and  $\sigma_{NLO}=2636$ fb respectively.
Varying the scale up and down by a factor 2  in a uniform way changes the
cross section by 70\% in the LO case, while in the NLO case we have
obtained a variation of the order 33\%. 

Subsequently, In Fig.~\ref{fig:scale} we show the result for the  scale
dependence graphically. 

In the next step the differential distributions are depicted  together with
their dynamical K-factors. Invariant mass and rapidity of the  two-$b$-jet
system are presented in Fig.~\ref{fig:mbb} and in  Fig.~\ref{fig:ybb}
respectively. Clearly, the distributions show the same large corrections,
which turn out to be relatively constant.

This large scale variation and the size of the corrections themselves, imply
that if a meaningful analysis were required in the present setup, additional 
cuts on extra jets must be introduced in order to reduce the NLO corrections.

\section{Summary}

A brief summary of the calculations of NLO QCD corrections 
to the   $pp\rightarrow
t\bar{t}b\bar{b} + X$ process at the LHC has been presented.   The QCD
corrections to the integrated cross sections are found to be very large,
changing the leading-order results by about 77\%.  The distributions show the
same large corrections. Moreover, the shapes  of some kinematical
distributions change appreciably compared to  leading order. The residual
scale uncertainties of the NLO predictions are at the 33\% level.

\vspace{1cm}

Work supported in part by the Initiative and Networking Fund of the Helmholtz
Association, contract HA-101 ("Physics at the Terascale") and by the RTN
European Programme MRTN-CT-2006-035505 HEPTOOLS - Tools and Precision
Calculations for physics Discoveries  at Colliders. 

\providecommand{\href}[2]{#2}\begingroup\raggedright

\end{document}